\documentclass{ws-procs975x65}

\def\beq{\begin{equation}}
\def\eeq{\end{equation}}

\begin{document}

\title{Phenomenology of Causal Dynamical Triangulations}

\author{Jakub Mielczarek}

\address{Institute of Physics, Jagiellonian University\\
{\L}ojasiewicza 11, 30-348 Cracow, Poland\\
$^*$E-mail: jakub.mielczarek@uj.edu.pl\\
www.jakubmielczarek.com}

\begin{abstract}
The four dimensional Causal Dynamical Triangulations (CDT) approach to 
quantum gravity is already more than ten years old theory with numerous 
unprecedented predictions such as non-trivial phase structure of gravitational 
field and dimensional running. Here, we discuss possible empirical consequences 
of CDT derived based on the two features of the approach mentioned above. 
A possibility of using both astrophysical and cosmological observations to test 
CDT is discussed. We show that scenarios which can be ruled out at the empirical 
level exist.
\end{abstract}


\bodymatter

\section{Introduction}

The characteristic feature of physical systems composed of a huge 
number of non-linearly coupled degrees of freedom is emergence 
of the so-called \emph{phases}, which correspond to different forms of 
internal organization. 

The gravitational field seem to fulfill the criteria for the non-trivial 
phase structure to occur. This is because the system is described 
by non-linear filed theory characterized by infinite number of degrees 
of freedom\footnote{In quantum version of the theory, the number 
of degrees of freedom theory is expected to be huge but finite.}. 

This presumption is materialized in the results obtained within 
Causal Dynamical Triangulations (CDT)\cite{Ambjorn:2012jv} 
approach, which aims to describe quantum nature of the gravitational 
interactions by employing path integral formulation of quantum 
mechanics. The most up-to-date studies of CDT predict existence of 
three phases of gravity, together with an additional \emph{bifurcation} 
sub-phase\cite{Ambjorn:2015qja}. The phases are separated 
by transition lines, among which first and second order phase 
transitions have already been detected\cite{Ambjorn:2011cg}. 

The quantity which is especially handy and useful in characterizing 
phases of gravity is the spectral dimension $d_S(\sigma)$, employing
a random walk process on the considered quantum space-time. 
In particular, in the ``most classical" phase $C$, the spectral dimension 
takes the value 4 for large diffusion time $\sigma$. However, at the 
short scales (small diffusion times) the value of $d_S$ decreases, 
which can be captured by the following parametrization:  
\begin{equation}
d_S(\sigma) =  4 -\frac{2-\epsilon}{1+\sigma E^2_*}.
\label{CDTSD2}
\end{equation}
Here, $E_*$ is a characteristic energy scale of the dimensional 
reduction and the value of $\epsilon$ depends on the location 
in the phase $C$ at the phase diagram.  The values of $\epsilon$ 
providedby the numerical simulations range form 
$\epsilon \approx 0 \ \ (d_S(0)\approx 2)$ \cite{Ambjorn:2005db}  
to $\epsilon \approx -1/2 \ \ (d_S(0)\approx 3/2)$ \cite{Coumbe:2014noa}.

\section{Modified dispersion relation}

The definition of spectral dimension is rooted in the diffusion 
process, which depends on spectra of Laplace operator defined 
on a given quantum manifold. As discussed in Ref.~\citenum{Mielczarek:2015cja}, 
(under certain assumptions) form of the Laplace operator and 
consequently a dispersion relation for massless particles 
can be reconstructed form the diffusion time dependence of the 
spectral dimension. In particular, assuming the dispersion relation 
in the form $E=\Omega( p )$, the following asymptotic behaviors 
of the $\Omega( p )$ function are obtained with use of Eq. (\ref{CDTSD2}):    
\begin{equation}
\Omega_{\rm IR} ( p ) \approx p+\frac{E_*}{15}(2-\epsilon) \left( \frac{p}{E_*}\right)^3   
\ \ \ \text{and} \ \ \  
\Omega_{\rm UV} ( p ) \approx  \frac{2}{3} E_* \left( \frac{p}{E_*}\right)^{3-3\epsilon}.   
\label{Lim}       
\end{equation}

The IR approximation can be applied to study propagation of 
high energy astrophysical photons. For example, using observational 
constraints on the energy-dependence of the group velocity $\left(v_{\rm gr} 
:= \frac{\partial \Omega ( p )}{\partial p}\right)$ of photons from the 
GRB 090510 source \cite{Vasileiou:2013vra} one can derive the 
following constraint on the energy scale of the dimensional reduction: 
$E_* >   6.7 \cdot 10^{10} {\rm GeV\  at}\ (95 \% {\rm CL})$. The 
``low-energy" (below around 10 EeV) dimensional reduction is, therefore, 
observationally excluded. 

Another possible application of the modified dispersion relation  
$E=\Omega( p )$ are cosmological perturbations in the early 
universe. At the phenomenological level, the dispersion relation 
can be introduced by replacing the momentum-space Laplace 
operator $\Delta_k$ contributing to the Hamiltonian of the type 
(details depend on which kind of the cosmological perturbations
is considered):  
\begin{equation}
H_{\phi} = \frac{1}{2} \int d^3{\bf k} \left\{  \frac{1}{a^2} \pi_{\bf k} \pi_{\bf -k} - \phi_{\bf k} 
\underbrace{\left[ - a^2 \Omega(k/a)^2\right]}_{=\Delta_k} \phi_{\bf -k}\right\}, 
\label{H}
\end{equation}
such that $\Delta_k \rightarrow -k^2$ in the classical limit and
$a$ denotes a scale factor. 

It is worth noticing that the method of introducing the effects of 
dimensional reduction applied here differs from the one 
considered in Ref.~\citenum{Amelino-Camelia:2013tla}. In that 
reference, modified dispersion relation has been introduced 
at the level of time-dependent speed of propagation. In our 
opinion, it is better justified to introduce the effect of modified 
dispersion relation at the level of the Fourier space representation 
of the Hamiltonian where the dispersion relation contributes explicitly.

\begin{figure}[h]
\begin{center}
$ \begin{array}{lr}   
a) \includegraphics[width=2.3in]{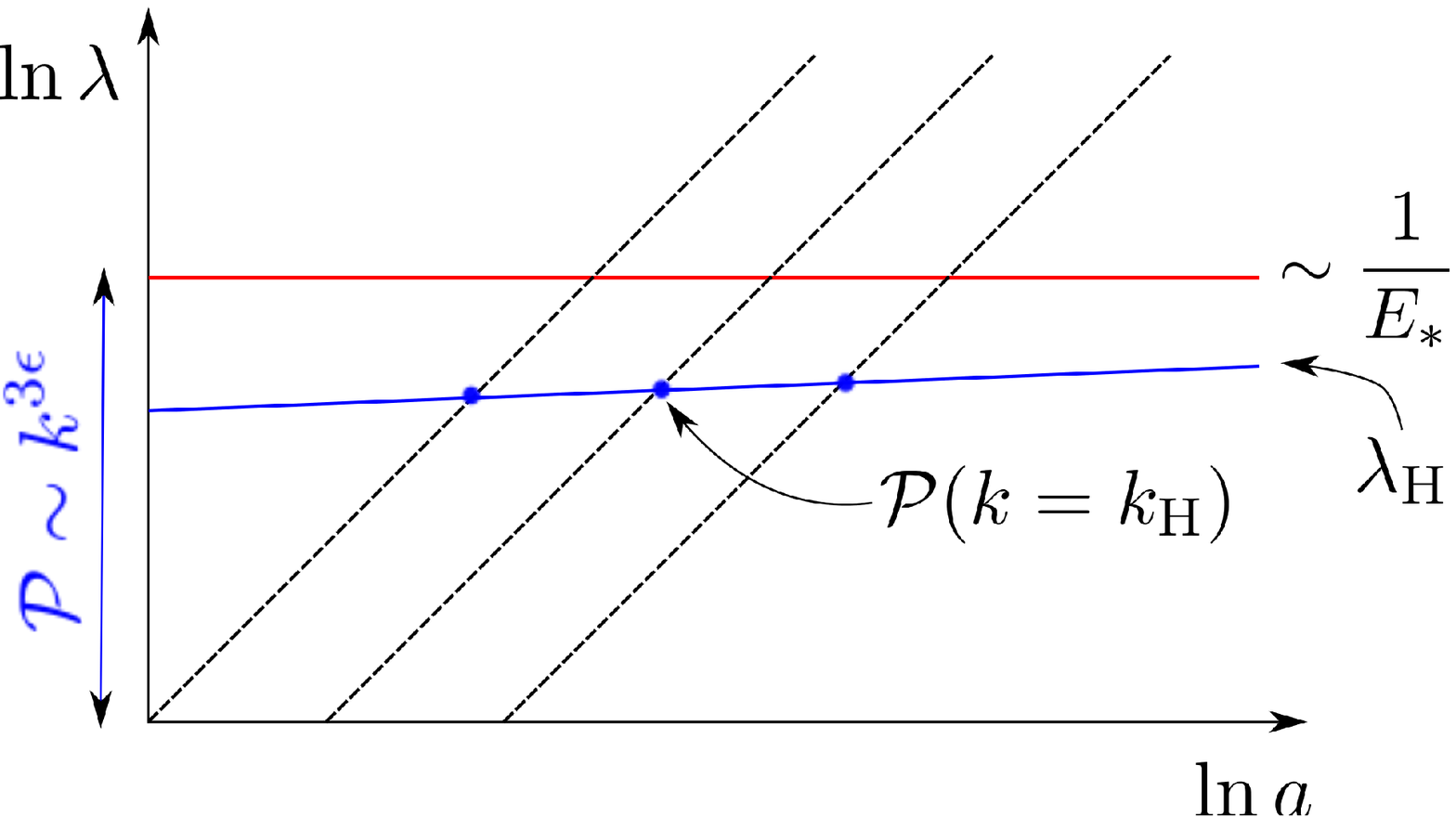} &
b) \includegraphics[width=2.3in]{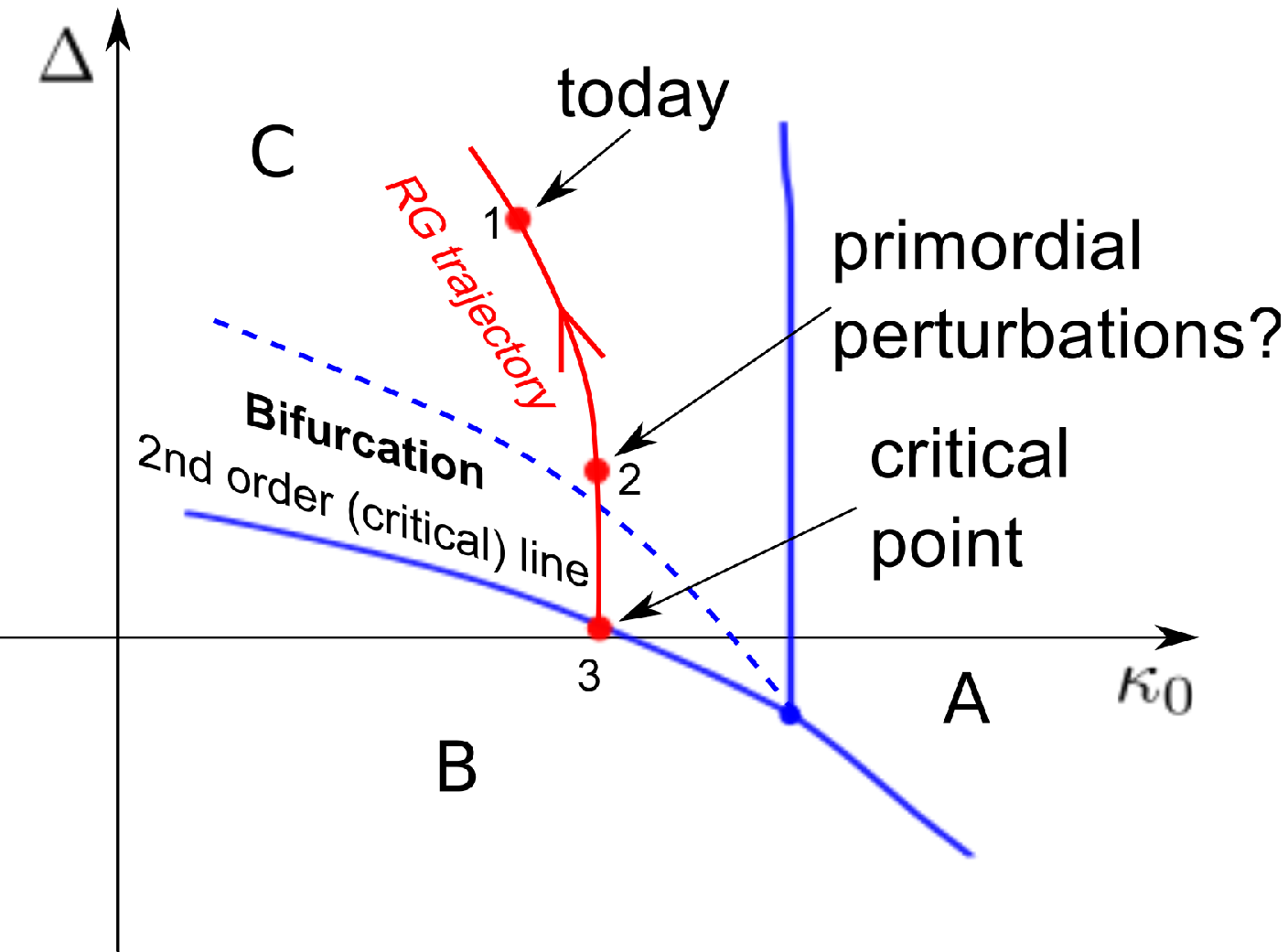} \\
\end{array} $
\end{center}
\caption{a) Red-shifting of physical length scales, Hubble radius and 
scale of the dimensional reduction. b) Phase diagram of CDT and speculated 
physical RG trajectory.}
\label{Fig}
\end{figure}

Analysis of the vacuum-normalized perturbations described by 
the Hamiltonian of the type (\ref{H}) leads to the following expression 
for the spectral index of the scalar perturbations:
\begin{equation}
n_S-1 \equiv \frac{d \ln \mathcal{P}(k=k_H) }{ d\ln k} \approx 
\frac{3\epsilon r}{r+48(\epsilon-1)}.
\label{nS}
\end{equation}
Here, $\mathcal{P}(k=k_H)$ is the amplitude of the scalar 
perturbations at the Hubble radius crossing (see Fig.~\ref{Fig}) 
and $r$ is the tensor-to-scalar ratio of the primordial perturbations. 

During derivation of Eq. (\ref{nS}) it has been assumed that the Hubble radius 
is smaller that the length scale of the dimensional reduction. As discussed 
in Ref.~\citenum{Mielczarek:2015cja}, this leads to predictions being in conflict with 
the up-to-date PLANCK and BICEP II data. If the case when the Hubble radius 
is much bigger than the scale of the dimensional reduction the classical results 
which (for certain type of inflationary potential) agree with the cosmological data 
are recovered. 

\section{Phase transitions and gravitational defects}

So far we considered points in the phase $C$ corresponding either 
to current state of the universe in case of the astrophysical 
constraints (point 1 in Fig.~\ref{Fig})  or to the state of the universe 
when the primordial cosmological perturbations were formed 
(point 2 in Fig.~\ref{Fig}). The two points are characterized 
by different energy scales (energy densities) and are connected  
by Renormalization Group (RG) trajectory realized in the observed
Universe. 

A worth considering possibility is that following backward the 
RG trajectory one ends up at the second order transition line 
(point 3 in Fig.~\ref{Fig}), which separates the bifurcation 
sub-phase of the phase $C$ from the non-geometric phase $B$. 
The phase transition, being an example of the \emph{geometrogensis} 
process, provides numerous prospects for building  
phenomenology of CDT. 

In particular, introduction of time scale to the $B-C$ phase transition lead us to 
the domain of non-equilibrium processes. If one would pass across the 
transition line infinitely slowly the system would have enough time to relax 
to a single global new ground state. However, passing trough the transition 
point in a finite amount of time does not allow to relax to a single ground 
state and collection of the so-called domains is formed (by virtue of the  
Kibble-Zurek mechanism). 
 
Some properties of the non-equilibrium transition might be estimated with use 
of the characteristics of the equilibrium phase transitions. In particular, typical sizes 
of the domains are $\xi\approx \xi_0 \left(\frac{\tau_Q}{\tau_0}\right)^{\frac{\nu}{\nu z+1}} 
\approx l_{\rm Pl}$, where $l_{\rm Pl}$ is the Planck length. In the case without  
cosmic inflation, the present size of the domains might be estimated as follows: 
$\xi_{\rm today} \approx \xi  \frac{T_{\rm Pl}}{T_{\rm CMB}} \sim 1\ {\rm mm}$.

At the boundaries separating the different ground states (domains) the gravitational 
defects are formed. With use of $\xi_{\rm today}$ an average defect concentration 
is $d \sim \frac{1}{\xi^3_{\rm today}} \sim \frac{1}{ {\rm mm}^3}$. Because any of the 
defects is observed, the gravitational version of the topological defect problem must 
be solved.  The solution might be provided by the phase of inflation, diluting the 
concentration of defects to the level beyond the observational threshold. On the other 
hand (for obvious reasons) presence of the cosmic inflation makes confrontation of 
the gravitational phase transitions with observations much more difficult. 
 
\section{Conclusions}

We have presented some possible paths allowing for construction of  
phenomenology of the CDT approach to quantum gravity. While at the 
moment none of the predictions of the CDT can be approved, some scenarios 
seem to be in conflict with the up-to-date observational data. The results 
encourage to take further attempts in Socratic debate with Nature on a 
role of CDT in description of gravitational phenomena at the Planck scale. 
 
\section*{Acknowledgments}

Author is supported by the Grant DEC-2014/13/D/ST2/01895 of the National Centre of
Science and the Iuventus Plus grant No.~0302/IP3/2015/73 from the Polish Ministry of 
Science and Higher Education.

\end{document}